\documentclass[aps,prl,amsmath,showpacs,floatfix,superscriptaddress]{revtex4}
\usepackage{bm}
\usepackage{amssymb}
\usepackage{graphicx}

\newcommand{\vc}[1]{\mbox{\boldmath{$#1$}}}

\begin{document}
\title{Diffusion and dispersion of passive tracers: Navier-Stokes versus MHD turbulence} 
\author{Wolf-Christian M\"uller}
\affiliation{Max-Planck-Institut f\"ur Plasmaphysik, 85748 Garching, Germany}
\email{Wolf.Mueller@ipp.mpg.de}
\author{Angela Busse}
\affiliation{Max-Planck-Institut f\"ur Plasmaphysik, 85748 Garching, Germany}
\affiliation{Universit\"at Bayreuth, Theoretische Physik II, 95440 Bayreuth, Germany}
\email{Angela.Busse@ipp.mpg.de}

\begin{abstract} 
A comparison of
turbulent diffusion and pair-dispersion in homogeneous, macroscopically isotropic 
Navier-Stokes (NS) and nonhelical magnetohydrodynamic (MHD) turbulence based on high-resolution
direct numerical simulations is presented.  Significant differences
between MHD and NS systems are observed in the pair-dispersion
properties, in particular a strong reduction of the separation
velocity in MHD turbulence as compared to the NS case.  It is
shown that in MHD turbulence the average pair-dispersion is slowed down 
for $\tau_\mathrm{d}\lesssim t\lesssim 10 \tau_\mathrm{d}$, $\tau_\mathrm{d}$ being the Kolmogorov time, due to the
alignment of the relative Lagrangian tracer velocity with the local magnetic field.  
Significant differences in
turbulent single-particle diffusion in NS and MHD turbulence are not
detected. The fluid particle trajectories in the vicinity
of the smallest dissipative structures are found to be
characterisically different although these comparably rare events have
a negligible influence on the statistics investigated in this work.
\end{abstract}
%
\maketitle 
The diffusive effect of turbulence on contaminants passively advected
by the flow is of great practical and fundamental interest.  While the
study of passive scalars \cite{warhaft:review} usually reverts to the
Eulerian description of the flow, the Lagrangian point of view has
proven to be very fruitful regarding investigations of turbulent
diffusion and pair-dispersion \cite{yeung:review, sawford:review} as
well as for the fundamental understanding of turbulence
\cite{falkovich_gawedzki_vergassola:review}.  The three-dimensional
dynamics of passive tracers in neutral fluids has been subject of
various experimental (for recent works see e.g.
\cite{voth_satyanarayan_bodenschatz:exp1,ott_mann:expdisp,mordant_pinton:strucexp})
and numerical, e.g. \cite{yeung_pope:singleparticle,yeung_borgas:dispersion1,biferale_boffetta_celani:bifreview,
ishihara_kaneda:dispsim},
investigations.  Related problems regarding the turbulent diffusion of
magnetic fields and the influence of turbulent magnetic fields on particle diffusion 
have been studied extensively in space and astrophysics, see e.g.
\cite{jokipii_parker:magdiff,jokipii_giacalone:cosrayturbmag,cattaneo:2Dturbdiff,maron_chandran_blackman:magdiff,matthaeus_etal:magdiff,
pommois_zimbardo_veltri:montecarlosolar}, as well as in the context of
magnetically confined nuclear-fusion plasmas, see for example
\cite{rechester_rosenbluth:fusiondiff,krommes_oberman_kleva:fusiondiff,isichenko:magdiff1}.
 
This Letter reports a first effort to identify differences in the
diffusion and dispersion properties of turbulent flows in electrically
conducting and in neutral media.  To this end the
dynamics of fluid particles is studied via high-resolution direct
numerical simulation of passive tracers immersed in fluids that are described
by the incompressible magnetohydrodynamic (MHD) and the Navier-Stokes
(NS) approximation.

Using the vorticity, $\vc{\omega}=\nabla\times\vc{v}$, and a uniform mass density, $\rho_0=1$,
the non-dimensional incompressible MHD equations are given by
\begin{eqnarray}
\partial_t\vc{\omega} &=& \nabla\times\left[\vc{v}\times\vc{\omega}
-\vc{b}\times(\nabla\times\vc{b})\right]+\mu\Delta\vc{\omega}\label{mhd1}\\
\partial_t\vc{b} &=&\nabla\times\left(\vc{v}\times\vc{b} \right)+\eta\Delta\vc{b}\label{mhd2}\\
\nabla\cdot\vc{v}&=&\nabla\cdot\vc{b}=0\label{mhd3}\,.
\end{eqnarray}
The dimensionless molecular diffusivities of momentum and magnetic
field are denoted by $\mu$ and $\eta$, respectively.  The magnetic
field \vc{b} is given in Alfv\'en-speed units. The Navier-Stokes
equations which govern the motion of an electrically neutral fluid are
obtained by setting $\vc{b}\equiv 0$ in
Eqs. (\ref{mhd1})-(\ref{mhd3}). 

The MHD/Navier-Stokes equations are solved by a standard
pseudospectral method in a triply periodic cube of linear extent
$2\pi$. The velocity field at the position \vc{X} of a tracer particle
is computed via tricubic polynomial interpolation and then used to move the
particle according to

\begin{equation}
\dot{\vc{X}}(t)\equiv\vc{V}(t)=\vc{v}_{{\mathrm{intpol}}}(\vc{X},t)\,.\label{diffeq}
\end{equation}
Eq. \ref{diffeq} is solved by a midpoint method which is
straightforwardly integrated into the leapfrog scheme
that advances the turbulent fields.
Test calculations using Fourier interpolated `exact' representations of turbulent velocity fields
have shown that the chosen tricubic polynomial interpolation delivers sufficient precision 
with a mean relative error at $512^3$ resolution of $\sim O(10^{-3})$. 
In addition tricubic interpolation is numerically much cheaper than the nonlocal spline approach 
(cf. \cite{yeung_pope:trackalgo}), especially on computing architectures with distributed memory. 
The initial particle positions are forming tetrads that are spatially
arranged to lie on a randomly deformed cubic super-grid with a maximum
perturbation of $25\%$ per super-grid cell.  This configuration
represents a compromise between statistical independence of particle
dynamics and a space-filling particle distribution
(cf. \cite{yeung_borgas:dispersion1,biferale_boffetta_celani:lagrapairs}). In
addition well-defined initial particle-pair separations $\Delta_0$
(cf. Table \ref{tab1}) are realized by the tetrad grouping.

Distances are given in units of the Kolmogorov dissipation length
$\ell_\mathrm{d}=(\mu^3/\varepsilon^\mathrm{K})^{1/4}$ defined with
the kinetic energy dissipation rate $\varepsilon^\mathrm{K}$ that is
part of the total dissipation rate
$\varepsilon=\varepsilon^\mathrm{K}+\varepsilon^\mathrm{M}=\int_V\mathrm{d}V(\mu\omega^2+\eta
(\nabla\times \vc{b})^2)$. In this work $k_\mathrm{max}\ell_\mathrm{d}\approx 1.6$. 
This fulfills the widely accepted
resolution criterion introduced in \cite{yeung_pope:singleparticle} and 
corresponds to a dissipative energy fall-off of 
about 4 decades in the dissipation range avoiding interpolation problems at grid-scales.  
Intervals of time are
given in units of the large-eddy turnover time, $T_0=\pi/(2E)^{1/2}$,
$E=E^K+E^M=\int_V\mathrm{d}V(v^2+b^2)/2$ being the total energy, or in
multiples of the Kolmogorov time,
$\tau_\mathrm{d}=(\ell_\mathrm{d}^2/\varepsilon^\mathrm{K})^{1/3}$, as
appropriate.

\begin{table}
\begin{center}
\begin{tabular}{c|c|c|c|c}
 group& $\Delta_{0}$ (NS) & $\Delta_{0}$ (MHD)  & particles & pairs \\   \hline    
 1&$1.8\ell_\mathrm{d}$ & $2.1\ell_\mathrm{d}$ & $4\cdot48^3$ & $331\,776$ \\ 
 2&$3.9\ell_\mathrm{d}$   & $4.6\ell_\mathrm{d}$ & $4\cdot48^3$ & $331\,776$ \\ 
 3&$7.9\ell_\mathrm{d}$   & $9.2\ell_\mathrm{d}$ & $4\cdot36^3$ & $139\,968$ \\ 
 4&$20\ell_\mathrm{d}$  & $23\ell_\mathrm{d}$& $4\cdot24^3$  & $41\,472$ \\  
 5&$98\ell_\mathrm{d}$ & $115\ell_\mathrm{d}$ & $4\cdot24^3$  & $41\,472$  \\ \hline
 total& ---       & ---       &$1\,181\,952$ &$886\,464$ \\
\end{tabular}
\end{center}
\caption{\label{tab1}Particle groups and respective initial pair separations $\Delta_0$}
\end{table}
The simulations are carried out using a resolution of $512^3$ collocation
points with aliasing errors being treated by spherical mode truncation
\cite{vincent_meneguzzi:simul}. Quasi-stationary turbulence is
generated by a forcing which freezes all modes in the sphere 
$k\leq k_f= 2$. The frozen modes which are taken from DNS decaying turbulence 
sustain the turbulence gently via nonlinear interactions. It has been checked
that this way of driving does not introduce significant anisotropy 
by regarding direction-dependent Eulerian two-point statistics. 

Starting with a set of random fluctuations of \vc{v} (and \vc{b} in
the MHD case) with zero mean the driven flows reach quasi-stationary states during which the
total energy $E$ shows fluctuations $\lesssim 10\%$ around unity and
$E^\mathrm{M}/E^\mathrm{K}\approx 2$ (MHD). In both simulations the
total energy dissipation rate $\varepsilon$ is quasi-constant at about
$0.24$ with $\varepsilon^\mathrm{M}\approx 0.15$ in the MHD case.  The
turbulent fields in the MHD system have negligible magnetic and cross
helicity. The macroscopic Reynolds numbers are dimensionally estimated
using $\mu$, $\eta$, $\varepsilon$, $E$, and the kinetic energy
$E^\mathrm{K}$ as
$\mathsf{Re}=(E^\mathrm{K})^{1/2}E^{3/2}/(\varepsilon \mu)$
(hydrodynamic) and $\mathsf{Rm}=\mathsf{Re}\mathsf{Pr_m}$ (magnetic)
with the magnetic Prandtl number $\mathsf{Pr_m}=\mu/\eta$ set to
unity. The respective numerical values of the parameters are
$\mu=8\times10^{-4}$, $\mathsf{Re}\approx 5400$ (NS) and
$\mu=\eta=5\times10^{-4}$, $\mathsf{Re}\approx 5200$ (MHD).
Cases with $\mathsf{Pr_m}\neq 1$ while interesting due to their importance in 
the context of turbulent dynamos (see e.g. \cite{ponty_pouquet:magprandtl,schekochihin_brandenburg:magprandtl}) 
are  beyond the scope of this paper and will be addressed in future work.

After the runs have reached macroscopic quasi-equilibrium the
trajectories of massless point particles marking the fluid are traced
over $13.5T_0$ (NS) and $11T_0$ (MHD) corresponding to
$500\tau_\mathrm{d}$ and $350\tau_\mathrm{d}$, respectively. The
initial particle positions are chosen in five groups of tetrads with
different particle-pair separations (cf. Table \ref{tab1}).

In statistically isotropic turbulence single tracer
particles are expected \cite{taylor:turbdiff} to show a diffusive
time dependence of the mean-square particle displacement
$\langle(\vc{X}(t)-\vc{X}_0)^2\rangle\sim t$, $\vc{X}_0=\vc{X}(0)$,
for $t\gg T_\mathrm{L}$ where $T_\mathrm{L}$
is the autocorrelation time of the Lagrangian velocity
$\vc{V}(t)$. Here, $T_\mathrm{L}\approx 16\tau_\mathrm{d}$ (NS) and
$T_\mathrm{L}\approx 15\tau_\mathrm{d}$ (MHD).  If $t\ll T_\mathrm{L}$
ballistic scaling is predicted,
$\langle(\vc{X}(t)-\vc{X}_0)^2\rangle\sim t^2$.

\begin{figure}
\centerline{\includegraphics[width=0.5\textwidth]{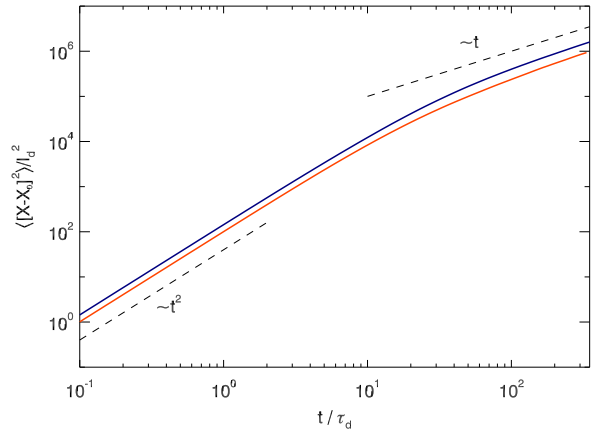}}
\caption{Evolution of normalized mean-square distance to initial position for turbulent single-particle 
diffusion 
in Navier-Stokes (black) and MHD turbulence.
The dashed lines indicate ballistic scaling $\sim t^2$ and diffusive behavior.}  
\label{f1} \end{figure}

In both simulations (cf. fig. \ref{f1}) ballistic scaling can be
identified up to about $T_\mathrm{L}$. Diffusive behavior is observed
for $t\gtrsim 50\tau_\mathrm{d}$.  At $t\gtrsim 70\tau_\mathrm{d}$ the
particles have traveled about $500\ell_\mathrm{d}$, i.e. half the size
of the simulation volume, and finite-size effects as well as the
influence of the large-scale driving can be detected. The normalized turbulent
diffusion coefficient $D_\mathrm{turb}(t_*)=\int_0^{t_*}\mathrm{d}\tau\langle
\vc{V}(t)\vc{V}(t+\tau)\rangle/\langle \vc{V}^2(t)\rangle$ with the avergage running over
all trajectories shows for both systems in the interval $0<t_*\lesssim 50 \tau_\mathrm{d}$ a steep increase 
with a subsequent saturation at the asymptotic value $T_\mathrm{L}$.

It is found that with regard to turbulent single-particle diffusion the NS and the MHD system show
no significant differences. The small offset
of the MHD displacement curve compared to the NS simulation is explained by the lower 
level of kinetic energy in the MHD system which is not fully compensated by the
applied normalization. An analytically predicted slowing down of diffusion (and dispersion) \cite{kim:diffu}
is not found here. The cited result is however based on the restricting assumption of a velocity field which is 
delta-correlated in time thereby neglecting the dynamically important adaptation of the velocity fluctuations 
to the magnetic field  structure.
    
The observed similarity of the curves in fig. \ref{f1} indicates
that statistics of single-particle 
trajectories is not a proper instrument to study the structural differences in the velocity field of the NS and MHD systems
caused by macroscopically isotropic magnetic field fluctuations
(cf. \cite{mueller_biskamp:3dmhdscale, biskamp_mueller:3dmhdscalprop,haugen_brandenburg:3dmhdsimfull} for numerical simulations). 

In this respect a more instructive diagnostic is relative
pair-separation or dispersion statistics where the separation,
$\vc{\Delta}=\vc{X}_1(t)-\vc{X}_2(t)$, of two particles is considered
\cite{batchelor:dispersion}.  The pair-separation in the ballistic
regime, $t\ll \tau_\mathrm{d}$, where the particle velocities are
finite and constant obeys the relation
$\langle(\vc{\Delta}(t)-\vc{\Delta}_0)^2\rangle\sim t^2$ (see
\cite{ott_mann:expdisp} for experimental and
\cite{yeung_borgas:dispersion1,biferale_boffetta_celani:bifreview} for
numerical results (all NS)) while for $t\gg T_\mathrm{L}$ diffusive
scaling holds, $\langle(\vc{\Delta}(t)-\vc{\Delta}_0)^2\rangle\sim t$
since the dynamics of particles forming a pair are statistically independent in this case.

\begin{figure}
\centerline{\includegraphics[width=0.5\textwidth]{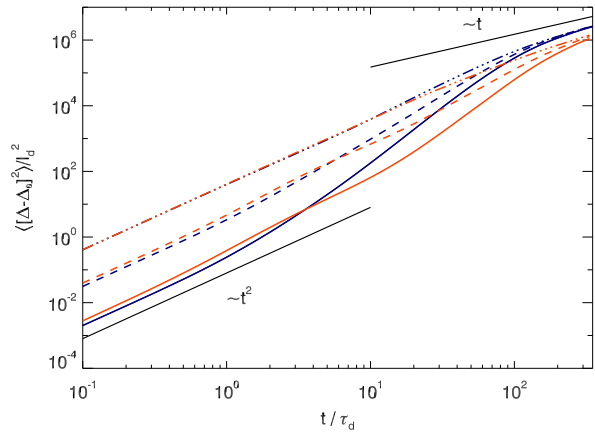}}
\caption{Evolution of normalized mean-square relative dispersion in Navier-Stokes (black) and MHD turbulence.
The behavior for three different initial pair-separations (particle groups in Table \ref{tab1}) 
are shown, solid: group 1,
dashed: group 3, dash-dotted: group 5. Straight lines denote ballistic $\sim t^2$ and diffusive  scaling.}
\label{f2} \end{figure}
The asymptotic limits can be identified in fig. \ref{f2} with
differences between NS and MHD statistics showing up at intermediate
times.  For clarity, only data based on the particle groups 1, 3, and
5 (cf. Table \ref{tab1}) are shown. In both systems the evolution of groups 2 and
4 is qualitatively similar to the behavior of group 3.
As expected ballistic scaling
$\sim t^2$ is visible in
both simulations for short times up to about $\tau_\mathrm{d}$. 
Eventually, for $t\gtrsim 160
\tau_\mathrm{d}$ an approach to the diffusive limit is seen
rudimentally.  The dispersion is accelerating during 
$10\lesssim t \lesssim 100$. However,  neither the Batchelor law
\cite{batchelor:dispersion}, $\langle(\vc{\Delta}-\vc{\Delta}_0)^2
\rangle=11/3C_2(\varepsilon\Delta_0)^{2/3}t^2$,
(cf. \cite{bourgoin_bodenschatz:dispexp}) nor Richardson scaling
\cite{richardson:dispersion},
$\langle\vc{\Delta}^2\rangle\sim\varepsilon t^{3}$,
(cf. \cite{ott_mann:expdisp, ishihara_kaneda:dispsim}) can be clearly
identified since the simulations do not generate sufficiently large inertial ranges.
In addition, the theoretical preconditions necessary for  Batchelor and Richardson behavior
are not satisfied in the simulations presented here. In particular, Batchelor dispersion requires
$\Delta_0$ to lie in the inertial range \cite{batchelor:dispersion} for recovering of the exact prefactor.
Richardson scaling is also not expected since it would entail very large $\Delta$ in the inertial range \cite{richardson:dispersion}. In addition Richardson behaviour would imply an appoach of the pair-separation curves to one 
universal scaling law independent of $\Delta_0$ which is not observed here.  

\begin{figure}
\centerline{\includegraphics[width=0.5\textwidth]{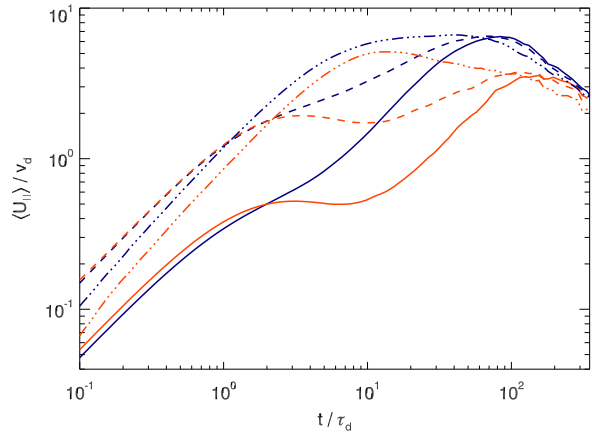}}
\caption{Normalized and averaged pair-separation velocity, $\langle U_\parallel\rangle$, in the direction of $\vc{\Delta}$ using the same symbols as in fig. 
\ref{f2}. Black lines: NS, grey lines: MHD}  \label{f3} \end{figure}
While the evolution of NS and MHD dispersion is qualitatively similar the acceleration phase in the 
MHD system is significantly 
delayed compared to
the NS case. The reason for this difference is found in the averaged separation velocity $\langle U_\parallel\rangle$
in the direction of the separation vector $\vc{\Delta}$ with
$\vc{U}=\dot{\vc{\Delta}}=U_\parallel\vc{\Delta}/\Delta+\vc{U}_\perp$ shown in fig. \ref{f3} in units of $v_\mathrm{d}=\ell_\mathrm{d}/\tau_\mathrm{d}$.  The separation velocity for all particle groups except
groups 5 (cf. Table \ref{tab1}) which have the largest initial
separation, $\Delta_0$, displays a continous increase before passing
through a slowing-down phase for $t\gtrsim \tau_\mathrm{d}$.  The
beginning of the slowing-down marks the point of time at which the
particles start to sense temporal fluctuations of the velocity field.
The subsequent acceleration of dispersion can be understood by the
increasing importance of sweeping by more coherent larger-scale
eddies.  The maximal separation velocity for all particle groups
except groups 5 is reached around $90\tau_\mathrm{d}$ (NS) and
$150\tau_\mathrm{d}$ (MHD).  There the mean pair separation is about
half the extent of the periodic simulation volume and the particles
start to approach each other again. The temporal shift of the MHD
maxima compared to the NS curves as seen in fig. \ref{f3} is explained by the smaller kinetic energy of the MHD
system and the observed stronger slowing-down of the average MHD pair-separation velocity.
The separation
velocity curves for the largest initial pair-separations (groups 5) do
not display the slowing-down phase since these pairs probe only the
largest spatial scales of the flow where the driving is governing
turbulent dynamics, cf. \cite{yeung_borgas:dispersion1}.

The main difference between the NS and MHD cases, however, lies in the
slowing-down phase which is much more pronounced in the MHD
simulation. 

The reason is the well-known anisotropy of turbulent eddies with
respect to the local magnetic field, see for example
\cite{shebalin_matthaeus:aniso,goldreich_sridhar:gs2,maron_goldreich:anisomhd,cho_vishniac:anisomhd,cho_lazarian_vishniac:anisomhd,
mueller_biskamp_grappin:anisomhd}.  Since small-scale eddies are
elongated in the local magnetic field direction MHD fluid elements are
more likely to travel in similar directions oriented along the local
magnetic field. The field-parallel velocity component is causing the effective particle pair-separation while the 
field-perpendicular components are associated with Alfv\'enic quasi-oscillations which do not contribute to 
the average particle displacement.  Contrary to the ballistic regime with
quasi-constant flow velocities for $t\ll\tau_\mathrm{d}$, the
anisotropy of the fluctuating velocity field at later times has a
constricting effect on turbulent dispersion compared to the NS case.
Consequently, the fluid particles are 
preferentially traveling along the magnetic lines of force which significantly reduces
dispersion in the field-perpendicular directions.

\begin{figure}
\centerline{\includegraphics[width=0.5\textwidth]{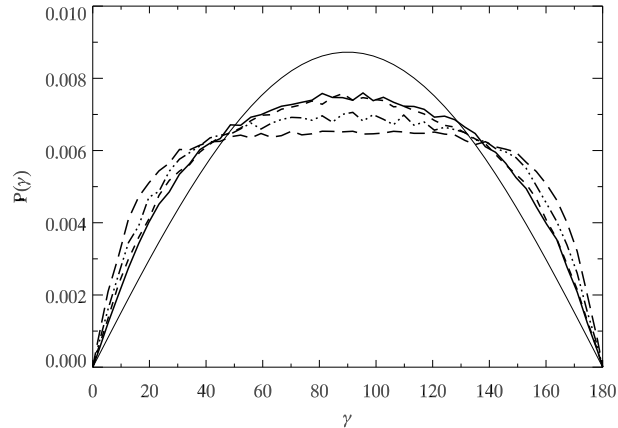}}
\caption{Probability distributions of the angle $\gamma$ between the relative velocity $\vc{U}=\dot\Delta$ of particles 
in group 1 (cf. Table \ref{tab1}) and 
the proxy $\overline{\vc{B}}=[\vc{b}(\vc{X}_1)+\vc{b}(\vc{X}_2)]/2$ at 
different times $t$ ($\tau_\mathrm{d}$: long-dash, $4\tau_\mathrm{d}$: dash-dot, $8\tau_\mathrm{d}$: short-dash,
$64\tau_\mathrm{d}$: solid). The thin line denotes the distribution $P(\gamma)=(\pi/360)\sin(\gamma)$ indicating 
isotropic dispersion.}  
\label{f4} \end{figure}
This conjecture is supported by fig. \ref{f4} which shows probability
distributions of the angle
$\gamma=\measuredangle(\vc{U},\overline{\vc{B}})$ for particle group 1
and points of time in the interval $0\leq t \leq 64\tau_\mathrm{d}$
introducing a rough proxy of the mean magnetic field at scale
$\Delta$, $\overline{\vc{B}}=[\vc{b}(\vc{X}_1)+\vc{b}(\vc{X}_2)]/2$.

For isotropic random velocities 
a sinusoidal distribution (thin solid line) would be
expected. However, it is seen that even for comparably large times the
distribution of the angle $P(\gamma)$ exhibits a clear deviation from
this behavior favoring velocities aligned with the
magnetic-field proxy $\overline{\vc{B}}$. The observed trend to
sinusoidality with increasing time is due to $\langle\Delta\rangle$
approaching the largest scales of the flow which leads to 
weakly correlated fluctuations in $\vc{U}$ and $\overline{\vc{B}}$.  This trend is
limited by the way of forcing chosen in this work.

Apart from pair-dispersion, tracer dynamics display another interesting difference between
Navier-Stokes and MHD turbulence. At smallest scales in the
vicinity of the most singular dissipative structures the tracer
trajectories differ significantly.  In the NS simulation where the
smallest flow structures are vortex filaments the fluid particles
describe helical motions around the filaments (cf. also \cite{biferale_boffetta_celani:bifreview}). In contrast, 
vorticity sheets typical for smallest-scales in MHD turbulence
lead to characteristic kinks in the tracer path. While the resulting
trajectories are strongly different and characteristic for the
respective turbulent system, these events occur too seldomly to have a
measurable effect on the statistics of diffusion and dispersion
regarded in this paper.
 
In summary it was shown by comparison of direct numerical simulations of macroscopically isotropic Navier-Stokes (NS) 
and nonhelical magnetohydrodynamic (MHD)
turbulence that the magnetic field in MHD turbulence slows down average pair-dispersion 
for intermediate times, $\tau_\mathrm{d}\lesssim t\lesssim 10 \tau_\mathrm{d}$,
compared to NS behavior. This effect is shown to be due to alignment of  
turbulent velocity and magnetic field fluctuations.
Significant differences in turbulent single-particle diffusion
could not be detected. Fluid particle trajectories in the vicinity of the strongly dissipative structures 
are characteristically different although these events have a negligible influence on the statistics investigated
in this work.
\acknowledgments
The authors would like to thank Holger Homann and Rainer Grauer (Ruhr-Universit\"at Bochum) for stimulating discussions and 
gratefully acknowledge support by Lorenz Kramer and Walter Zimmermann (Universit\"at Bayreuth).
Computations were performed on the Altix 3700 system at the Leibniz-Rechenzentrum, Munich. 
\newcommand{\nop}[1]{}

\end{document}